\documentclass[twocolumn,showpacs,amsmath,amssymb,prl,superscriptaddress]{revtex4}

\usepackage{graphicx}
\usepackage{dcolumn}
\usepackage{bm}
\usepackage{amssymb}
\usepackage{bbm}
\usepackage{calc}

\begin{document}

\title{Mid-Infrared Optical Frequency Combs based on Crystalline Microresonators}

\author{C.~Y.~Wang}
\altaffiliation{There authors contributed equally to this work.}
\affiliation{Max-Planck Institut f\"ur Quantenoptik, Hans-Kopfermann Strasse 1, D-85748 Garching, Germany}
\affiliation{\'{E}cole Polytechnique F\'{e}d\'{e}rale de Lausanne (EPFL), CH-1015 Lausanne, Switzerland}
\affiliation{Menlo Systems GmbH, Am Klopferspitz 19a, D-82152 Martinsried, Germany}

\author{T.~Herr}
\altaffiliation{There authors contributed equally to this work.}
\affiliation{Max-Planck Institut f\"ur Quantenoptik, Hans-Kopfermann Strasse 1, D-85748 Garching, Germany}
\affiliation{\'{E}cole Polytechnique F\'{e}d\'{e}rale de Lausanne (EPFL), CH-1015 Lausanne, Switzerland}

\author{P.~Del'Haye}
\altaffiliation{Current affiliation: National Institute of Standards and Technology, 325 Broadway, Boulder CO 80305, USA}
\affiliation{Max-Planck Institut f\"ur Quantenoptik, Hans-Kopfermann Strasse 1, D-85748 Garching, Germany}
\affiliation{Menlo Systems GmbH, Am Klopferspitz 19a, D-82152 Martinsried, Germany}

\author{A.~Schliesser}
\affiliation{Max-Planck Institut f\"ur Quantenoptik, Hans-Kopfermann Strasse 1, D-85748 Garching, Germany}
\affiliation{\'{E}cole Polytechnique F\'{e}d\'{e}rale de Lausanne (EPFL), CH-1015 Lausanne, Switzerland}

\author{J.~Hofer}
\altaffiliation{Current affiliation: Paul Scherrer Institute, 5232 Villigen PSI, Switzerland}
\affiliation{Max-Planck Institut f\"ur Quantenoptik, Hans-Kopfermann Strasse 1, D-85748 Garching, Germany}

\author{R.~Holzwarth}
\affiliation{Max-Planck Institut f\"ur Quantenoptik, Hans-Kopfermann Strasse 1, D-85748 Garching, Germany}
\affiliation{Menlo Systems GmbH, Am Klopferspitz 19a, D-82152 Martinsried, Germany} 

\author{T.~W.~H\"ansch}
\affiliation{Max-Planck Institut f\"ur Quantenoptik, Hans-Kopfermann Strasse 1, D-85748 Garching, Germany}
\affiliation{Ludwig-Maximilians-Universit\"at M\"unchen, Fakult\"at f\"ur Physik, Schellingstrasse 4/III, 80799 M\"unchen, Germany}

\author{N.~Picqu\'{e}}
\email{nathalie.picque@mpq.mpg.de}
\affiliation{Max-Planck Institut f\"ur Quantenoptik, Hans-Kopfermann Strasse 1, D-85748 Garching, Germany}
\affiliation{Ludwig-Maximilians-Universit\"at M\"unchen, Fakult\"at f\"ur Physik, Schellingstrasse 4/III, 80799 M\"unchen, Germany}
\affiliation{Institut des Sciences Mol\'{e}culaires d'Orsay, CNRS, B\^atiment 350, Universit\'{e} Paris-Sud, 91405 Orsay, France}

\author{T.~J.~Kippenberg}
\email{tobias.kippenberg@epfl.ch}
\affiliation{Max-Planck Institut f\"ur Quantenoptik, Hans-Kopfermann Strasse 1, D-85748 Garching, Germany}
\affiliation{\'{E}cole Polytechnique F\'{e}d\'{e}rale de Lausanne (EPFL), CH-1015 Lausanne, Switzerland}

\begin{abstract}

The mid-infrared spectral range ($\lambda \sim 2 \:\mu$m to $20 \:\mu $m) is of particular importance for chemistry, biology and physics as many molecules exhibit strong ro-vibrational fingerprints. Frequency combs \cite{Udem, Cundiff} - broad spectral bandwidth coherent light sources consisting of equally spaced sharp lines - are creating new opportunities for advanced spectroscopy \cite{Diddams, Thorpe, Mandon, Coddington, Bernhardt, Schliesser}. Mid-infrared frequency comb sources have recently emerged \cite{Adler, Leindecker, Maddaloni} but are still facing technological challenges, like achieving high power per comb line and tens of GHz line spacing as required for e.g. direct comb spectroscopy \cite{Diddams}.  Here we demonstrate a novel approach to create such a frequency comb via four-wave mixing in a continuous-wave pumped ultra-high Q crystalline microresonator made of magnesium fluoride. Careful choice of the resonator material and design made it possible to generate a broad comb of narrow lines in the mid-infrared: a vast cascade of about 100 lines spaced by 100 GHz spanning 200~nm ($\sim 10$~THz) at $\lambda=2.5 \:\mu$m. With its distinguishing features of compactness, efficient conversion, large mode spacing and high power per comb line, this novel frequency comb source holds promise for new approaches to molecular spectroscopy even deeper in the mid-infrared.

\end{abstract}

\maketitle

 Due to its scientific and technological significance, laser technology in the mid-infrared is an active area of research and development. The advent of a compact and versatile coherent light source in this region came after the invention of Quantum Cascade Lasers (QCLs) in 1994 \cite{Faist}. However, QCLs are intrinsically difficult to be passively mode-locked \cite{Wang1} and only active mode-locking has been unequivocally demonstrated creating a limited "comb" of ca. 0.3 THz bandwidth \cite{Wang2}.  Today, the most common approach to create frequency combs in the mid-infrared is to frequency down-convert a near-infrared comb through nonlinear processes, such as optical parametric oscillation \cite{Adler, Leindecker} or difference frequency generation \cite{Maddaloni}.

Here we demonstrate a novel route to frequency comb generation in the mid-infrared based on ultra high-Q crystalline optical microresonators \cite{KippenbergReview}. The particular advantages of microresonator-based frequency combs are the compact form factor, direct fibre optical coupling, high repetition rate and high power per comb line. It has also been shown that dispersion engineered microresonators enable the generation of spectra that cover more than one full octave \cite{DelHaye3, Okawachi}, as required for $f-2f$ phase stabilization. The underlying mechanism is cascaded four-wave mixing (FWM) caused by the third-order Kerr nonlinearity in high-Q whispering-gallery mode (WGM) microresonators, which was first demonstrated in silica microtoroids in the near-infrared \cite{Kippenberg, DelHaye1}. In this energy conserving process, two pump photons (frequency $\nu_{\mathrm{P}}$) are annihilated to give rise to a pair of signal and idler photons, whose frequencies are up-shifted ($\nu_{\mathrm{S}}$) and down-shifted ($\nu_{\mathrm{I}}$) such that ($2\nu_{\mathrm{P}} = \nu_{\mathrm{S}}+\nu_{\mathrm{I}}$). This process can cascade and thereby lead to the formation of an equidistant optical frequency comb ("Kerr" comb), whose spectral properties (offset frequency and mode spacing) can be stabilized \cite{DelHaye2}. Several microresonator platforms based on this mechanism have demonstrated Kerr frequency comb generation in the near-infrared region \cite{KippenbergReview, DelHaye1, Levy, Foster, Savchenkov, Razzari}. However, to move Kerr combs into the mid-infrared, one has to carefully consider many material properties such as transparency and dispersion, and most of the existing platforms cannot be used in the mid-infrared.

\begin{figure}[ht] 
 \begin{center}
 \includegraphics[width=8cm]{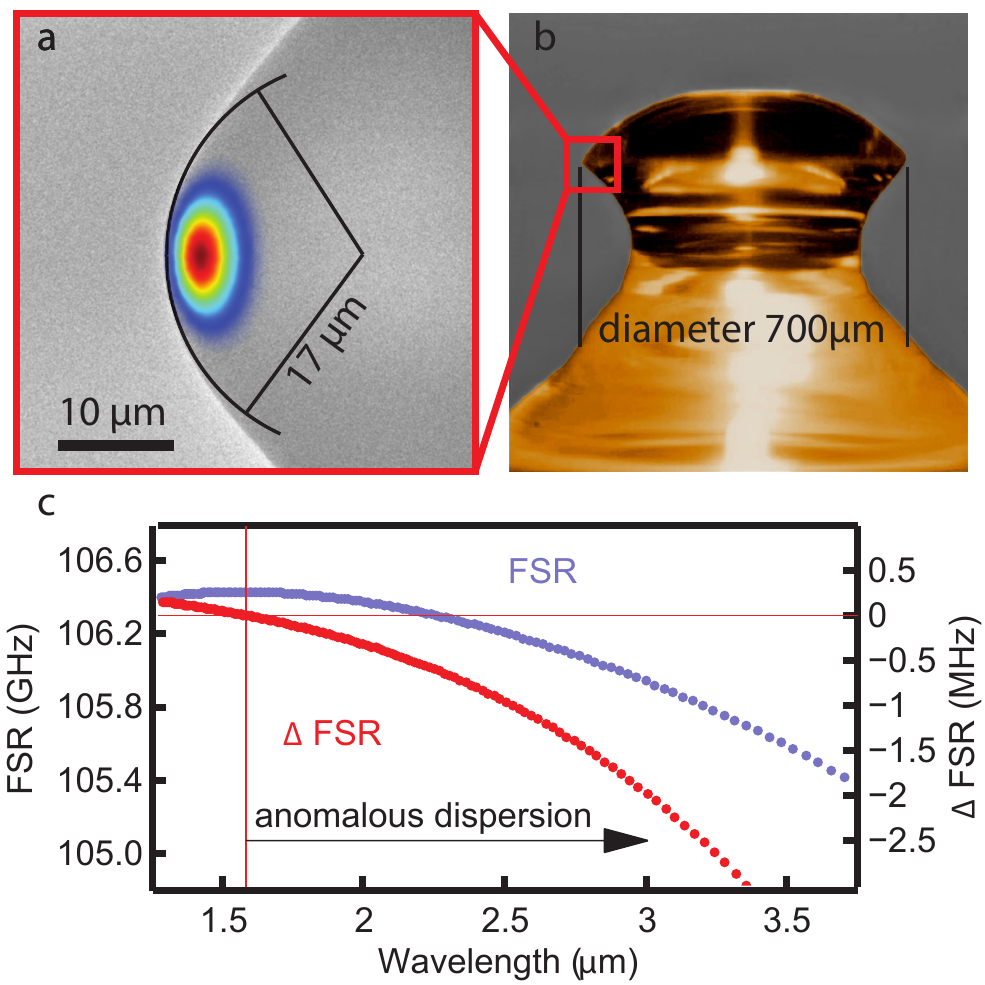}
 \end{center}
 \caption{\textbf{Properties of a $\mathrm{MgF_2}$ crystalline whispering-gallery mode microresonator.} (a) Optical microscope image of the $700$-$\mu$m diameter resonator used in this work, fabricated by polishing and shaping a UV grade $\mathrm{MgF_2}$ cylinder blank (b) Finite element simulation of the optical intensity profile of a whispering-gallery mode superimposed on a scanning electron microscope image of the resonator at wavelength $2.45\:\mu$m. The radius of curvature of the resonator protrusion, which confines the whispering-gallery mode in the azimuthal direction, is $17\:\mu$m. (c) The figure shows the simulated free spectral range for fundamental modes with mode number m ($FSR_m=(\nu_{m+1}-\nu_{m-1})/2$, blue curve) and the difference of the free spectral range between adjacent modes ($\Delta FSR_m=(FSR_{m+1}-FSR_{m-1})/2$, red curve) as a function of the wavelength for the resonator in (a), exhibiting an anomalous group velocity dispersion (GVD) above $\lambda=1.6 \:\mu$m.}
 \end{figure}
			
	For mid-infrared frequency comb generation, we developed ultra-high Q resonators made of crystalline magnesium fluoride ($\mathrm{MgF_2}$) as shown in Figure 1a,b following the approach first introduced in Ref.~\cite{Ilchenko}. The reason for choosing $\mathrm{MgF_2}$ is threefold: First, the transparency window of crystalline materials (such as $\mathrm{CaF_2}$ and $\mathrm{MgF_2}$) extends from the UV (ca. $160$~nm) to the mid-infrared (ca.~7~$\mu$m), enabling to achieve ultra-high Q ($>10^8$) in the mid-infrared-in stark contrast to fused silica or quartz, which always contain water and suffer from strong absorption above 2.2 $\mu$m. Second, the ability to generate optical frequency combs via cascaded FWM in the presence of self- and cross-phase modulation of the pump requires the group velocity dispersion (GVD) in the spectral region of interest to be anomalous \cite{Kippenberg, Matsko}, i.e. a free spectral range (FSR) of the cavity that reduces with increasing wavelength. Indeed, recent work \cite{Herr} shows that a certain minimal amount of anomalous dispersion is critical for the primary oscillating comb modes to be close to the pump thereby achieving low phase noise operation. We assess the dispersion properties by carrying out fully vectorial finite-element simulations \cite{Oxborrow, DelHaye4} including material and geometrical effects. Figure 1c shows the dispersion properties of a $700$-$\mu$m-diameter $\mathrm{MgF_2}$ crystalline resonator, revealing a GVD that is anomalous over the full mid-infrared transparency range. The dispersion requirement renders other materials, such as $\mathrm{Si_3N_4}$ microring resonators which are used successfully at shorter wavelength for Kerr-comb generation, unfavourable as their GVD is normal in the mid-infrared despite being transparent \cite{Okawachi}.  Third, the sign of the temperature coefficient of the refractive index $(dn/dT)$ and the thermal expansion coefficient $\alpha$ are both positive in $\mathrm{MgF_2}$ (for temperatures close to room temperature) which allows the cavity mode to be thermally self-locked to the pump laser frequency \cite{Carmon} and, in principle, enables the stabilization of the repetition rate and offset frequency of microresonator-based frequency combs \cite{DelHaye2}. On the contrary, materials with opposite signs of $dn/dT$ and $\alpha$ such as $\mathrm{CaF_2}$ do not allow for thermal self-locking, which results in large thermal oscillatory instabilities and the requirement for active locking scheme (or injection locking scheme).

\begin{figure*}[ht] 
 \begin{center}
 \includegraphics[width=15cm]{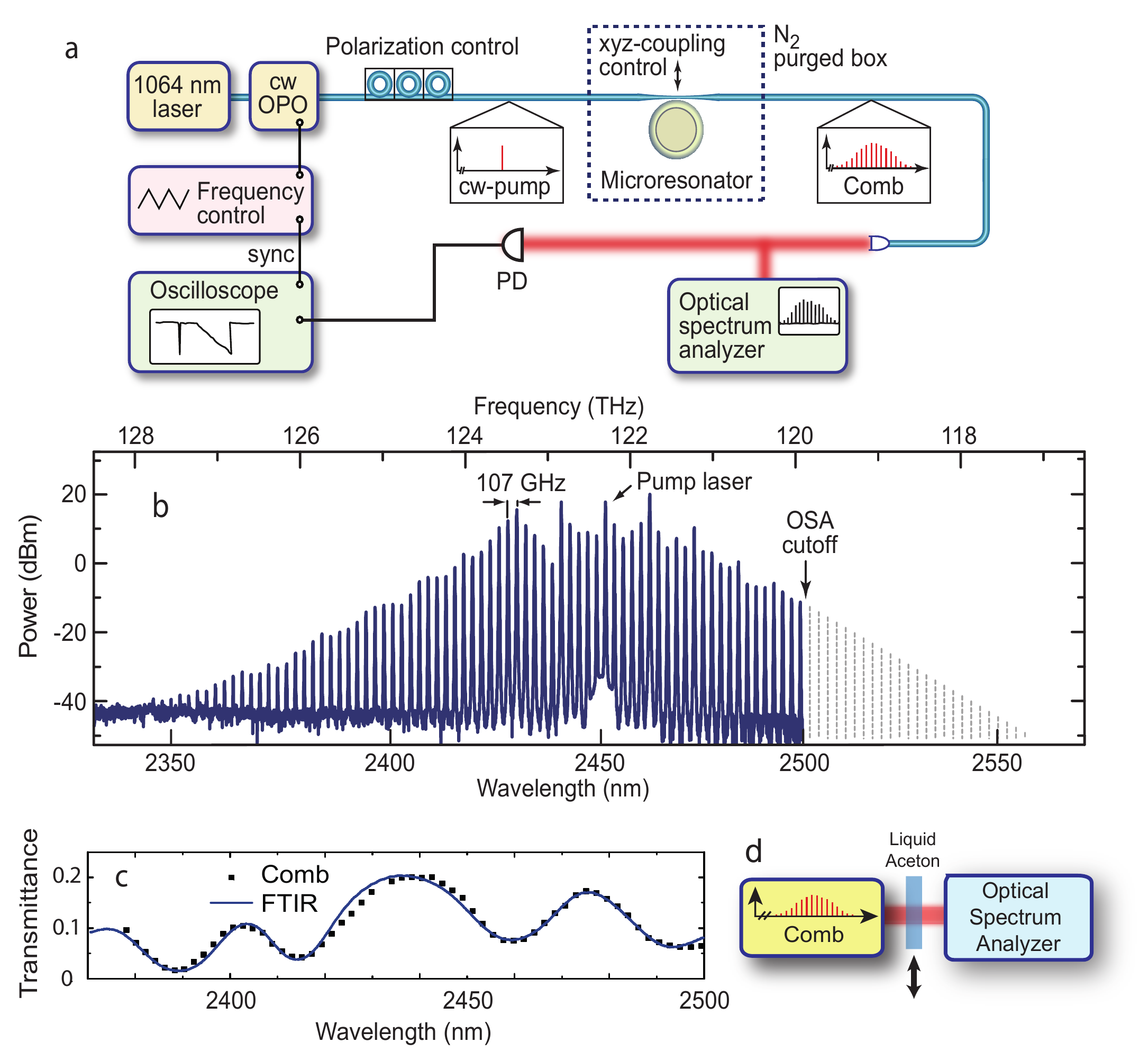}
 \end{center}
 \caption{\textbf{Mid-infrared optical frequency comb generation from a crystalline $\mathrm{MgF_2}$ microresonator.}  (a) The experimental setup consists of a continuous wave mid- infrared optical parametric oscillator that serves as the pump laser. The pump laser is coupled via a tapered fibre to a crystalline MgF2 microresonator. The generated frequency comb is detected using an optical spectrum analyser (OSA) with a cut-off wavelength of $2.5\:\mu$m (PD: photodetector). (b) The frequency comb spectrum recorded by the optical spectrum analyzer (OSA) around $\lambda=2.45\:\mu$m with a line spacing of 107 GHz generated from pumping a $700$-$\mu$m-diameter $\mathrm{MgF_2}$ resonator. The grey lines denote frequency components which are expected to exist based on a symmetric comb around the pump. (c) Proof-of-principle mid-infrared Kerr comb absorption spectroscopy experiment. The figure shows the transmittance of $0.5$ mm-thick liquid acetone (single-pass) recorded as the relative intensities of the attenuated and unattenuated Kerr-comb (shown in b) line intensities (dot) and independently by a Fourier-transform spectrometer (line). (d) Schematics of the acetone absorption experiment. }
 \end{figure*}

	        To fabricate the resonators, single-crystal excimer-grade $\mathrm{MgF_2}$ was first cut into cylinder blanks with several millimetre dimensions.  The resonators were then shaped and polished on an air-bearing spindle by diamond abrasives to achieve a smooth protrusion which provides an azimuthal WGM confinement, following the methods described in Ref.~\cite{Hofer}. A scanning electron microscope was used to measure the transverse radius of curvature of the resonator to be $17\:\mu$m (cf. Figure 1b). The resulting mode area of the WGM was determined in a fully vectorial finite element simulation. Superimposed on Figure 1 b is the intensity profile of the fundamental mode at $\lambda=2.45\:\mu$m. From these simulations we obtained an effective mode area $A_{\mathrm{eff}}=60~\mathrm{\mu m^2}$ and determined the effective nonlinearity of the resonator $\gamma_{\mathrm{eff}}=2 \pi n_2/\lambda A_{\mathrm{eff}}=4.3 \times 10^{-4}\: \mathrm{m^{-1} W^{-1}}$, where $n_2$ is the Kerr nonlinearity of $\mathrm{MgF_2}$ ($1 \times 10^{-20} \:\mathrm{m^2/W}$). The diameter of the resonators typically ranges from $500~\mu$m to $5$~mm, corresponding to a native resonator mode spacing in the range of 10-110 GHz.  Tapered optical fibres made of silica were used to characterize the optical properties of the resonators at $1550$~nm. This method \cite{Hofer} is particularly suitable as the refractive index of $\mathrm{MgF_2}$ ($n=1.37$) is close to that of silica, and critical coupling can be achieved. For pre-characterization, the quality factor of the resonances was measured with a tuneable, narrow-linewidth ($\Delta\nu \sim 3\:$kHz) single-mode fibre laser at $\lambda = 1550$~nm, yielding intrinsic Q-factors exceeding $10^9$, corresponding to an optical cavity finesse of $F=10^5-10^6$. Note that this quality factor is still far lower than the material limit of the ultra-pure crystalline magnesium fluoride. Additional polishing and cleaning steps of the resonators can reduce the scattering loss and even further increase the Q factors \cite{Savchenkov2}.

        We pumped the resonators by a continuous-wave (CW) mid-infrared laser based on an optical parametric oscillator that is tuneable between $2.4$ and $2.5 \:\mu$m with short term linewidth $<100$ kHz. Details of the laser source and the experimental setup can be found in the Methods section and in Figure 2a.  Optical power levels of $200~$mW to $1~$W were coupled into the resonators by employing a tapered fibre made of low-OH fused silica ($\sim 3.4~\mathrm{dB/m}$ at $2.5~\mu$m). The resonator and tapered fibre are embedded in a dry nitrogen purged environment, as water vapour is highly absorptive in the mid-infrared (prior exposure to air however does not result in accumulation of water layers as $\mathrm{MgF_2}$ crystals are hydrophobic). The output spectra were recorded by an optical spectrum analyzer (OSA) with cut-off wavelength at $2.5\:\mu$m. Figure 2b shows the optical frequency comb spectrum derived from the $700-\mu$m-diameter resonator pumped with ca. $600\:$mW of laser power at $\lambda=2.45 \:\mu$m. We observed more than 100 modes spaced by 107~GHz (corresponding to a span of over 200~nm), which corresponds to the free spectral range of the resonator ($FSR=c/2\pi R n_{\mathrm{eff}})$, where R is the cavity radius, c is the speed of light in vacuum and $\mathrm{n_\mathrm{{eff}}}$ is the effective refractive index. Due to the high optical finesse ($F=1.2 \times 10^5$) the circulating power in the resonator is approximately $P=1.1 \times 10^4 ~\mathrm{W}$. As the resonator has a third order nonlinearity, sidebands can be generated once the parametric threshold is exceeded.  Neglecting cavity dispersion the threshold for parametric oscillation is reached \cite{Kippenberg} when the Kerr nonlinearity induced frequency shift $\Delta\omega_{\mathrm{kerr}}$ equals half the cavity decay rate (the bi-stability point), i.e. when $\Delta\omega_{\mathrm{kerr}}=n_{\mathrm{2}}/n \cdot \omega \cdot (FP_{\mathrm{coupled}})/A_{\mathrm{eff}} =\kappa/2$, where $n$ is the refractive index,  $n_2$  the Kerr nonlinearity, $P_{\mathrm{coupled}}$ the power coupled into the resonator, $A_{\mathrm{eff}}$ the effective modal area, $F$ the cavity finesse, $\kappa$  the cavity decay rate and $\omega$ the optical angular frequency.  The estimated threshold for sideband generation is $P=4$~mW, which matches the experimental observation. The power per comb line ranges from micro-Watts up to several milli-Watts, significantly higher than the current state-of-the-art down-converted mid-infrared combs \cite{Adler, Leindecker}.  

       Frequency combs have recently demonstrated potential for advances in molecular spectroscopy \cite{Bernhardt2, Adler2}. To demonstrate that the high power per comb line and the wide mode spacing is well suited for the recording of broadband vibrational spectra in the liquid or solid phase, we performed a proof-of-concept absorption spectroscopy of $0.5$ mm-thick layer of liquid acetone using the comb in Figure 2b as light source and recording the attenuated spectra with the OSA. The result is shown in Figure 2c and validated by an independent measurement of the same acetone sample employing a conventional white light source and a Fourier transform spectrometer. In future experiments employing a mid-infrared disperser and detector array, we expect microresonator-based combs to be suitable light sources for direct molecular fingerprinting \cite{Diddams} which would vastly improve the acquisition time and sensitivity. Therefore, new opportunities for real-time spectroscopic investigation and optimization of chemical reactions may come into reach, including industrial real-time process control of e.g. pharmaceutical products.

\begin{figure*}[ht] 
 \begin{center}
 \includegraphics[width=15cm]{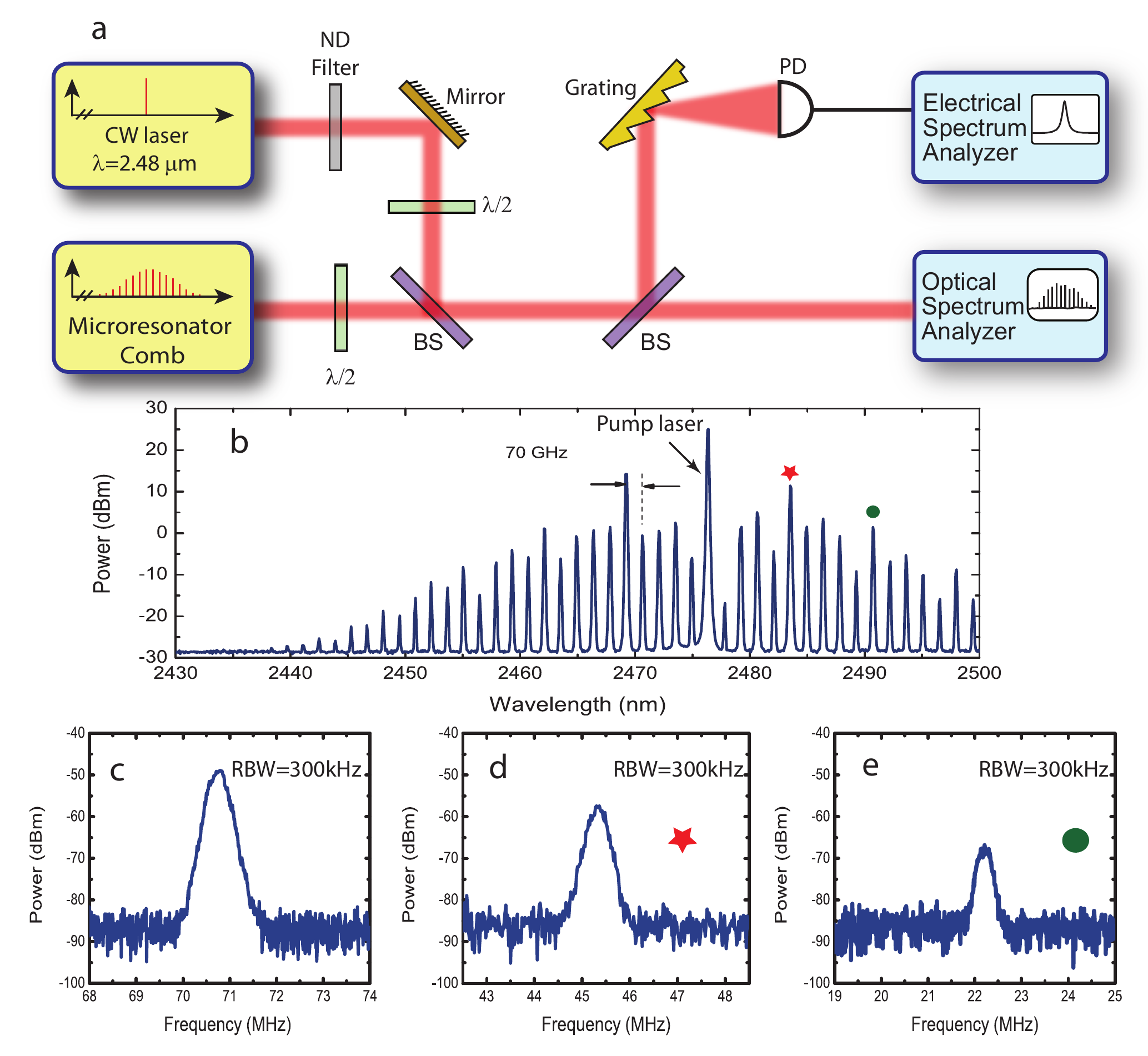}
 \end{center}
 \caption{\textbf{Phase noise investigation of the mid-infrared frequency comb modes.} (a) Experimental setup of the continuous-wave (CW) laser beat note measurement. BS: beam splitter, PD: photodetector. (b) Frequency comb spectrum with a line spacing of $70$ GHz used for the beat note measurement.  (c) shows the beat note between the CW laser and the pump, while (d) and (e) show the beat note between the CW laser and the comb modes (marked by the red star and the green dot in (b), respectively). All beat notes have a FWHM of approximately $300$ kHz, corresponding to the resolution bandwidth used in the measurement. No other peaks are observed in the measurement with a larger frequency span. Beat notes between the CW laser and other comb modes show the same linewidth.}
 \end{figure*}

        A key requirement of an optical frequency comb is low phase noise. We investigated the comb phase noise by detecting the beat notes between comb lines and an additional narrow-linewidth CW laser similar to the pump source. The experimental setup and results of the beat note measurement are shown in Figure 3. The comb spectrum (derived from a $1$-mm-diameter resonator) is shown in Figure 3b. After measuring the beat of the CW laser with the narrow-linewidth pump (Fig. 3c), the beat notes between the CW laser and different comb modes are recorded (Fig. 3d, e). All beat notes exhibit the same FWHM corresponding to the employed resolution bandwidth of $300~$kHz. Importantly, neither additional linewidth broadening of the comb modes relative to the pump nor multiple beat notes (in contrast to measurements on near-infrared crystalline, silicon nitride \cite{Herr} and fused quartz \cite{Papp} resonators, and in toroids generating broadband spectra \cite{DelHaye3}) are observed in this measurement, indicating that the comb lines exhibit a similar levels of phase noise as the CW pump laser.

	While parametric sidebands can be generated with the pump laser being close to the zero dispersion wavelength, a certain amount of anomalous dispersion is required to achieve low phase noise in Kerr combs, due to different possible four-wave mixing pathways as detailed in Ref.~\cite{Herr}. For low phase noise operation it is important to achieve a low ratio of $\kappa /|\Delta FSR|$, where $\kappa$ denotes the cavity decay rate and $\Delta FSR$ the variation of free-spectral range in the anomalous dispersion regime. In contrast to previous work in crystalline resonators, the present experiments operate in the strongly anomalous dispersion regime. This can be evidenced in the simulation in Figure 1c, which shows that the magnitude of $\delta FSR$ is increasing with wavelength monotonically above ca. $1.55\:\mu$m in $\mathrm{MgF_2}$ resonators (anomalous dispersion). Indeed, the observation that Kerr combs with low phase noise at $2.45\:\mu$m can be generated (while phase noise appears by pumping the same resonators at $1.55\:\mu$m) agrees with the theoretical prediction \cite{Herr}. Consequently, low phase noise Kerr combs can therefore be expected across the entire mid-infrared region. In the present work, the choice of pump wavelength of $\lambda = 2.45\:\mu$m originates from the current limitation of the available optical spectrum analyzer and the strongly increasing coupling fibre absorption beyond $2.5\:\mu$m. While the first reason is not a limiting factor of the system itself, the fibre absorption may be circumvented by employing e.g. a coupling prism \cite{Gorodetsky} or tapered chalcogenide fibres \cite{Magi} instead. 
	 
        Pertaining to the advantages of crystalline microresonator based mid-infrared frequency comb generators, it is noted that the native mode spacing in the range of $10-110$ GHz is a unique feature. On the one hand, the comb repetition rate and carrier envelope offset frequencies are accessible with fast photodetectors, electronics and digital signal processing; on the other hand, the comb modes may be individually accessed and even controlled. Indeed, a growing number of emerging applications require combs with large line spacing, such as astronomical spectrograph calibration, arbitrary optical waveform synthesis and spectroscopy. Line-by-line pulse shaping of individual comb lines \cite{Ferdous} may prove beneficial to coherent control of chemical reactions. 
        
        In summary, a crystalline microresonator-based optical frequency comb in the mid-infrared "molecular fingerprint" region is demonstrated for the first time. The high power per comb line and the potential to extend the comb to longer wavelengths make it a promising candidate for spectroscopic applications. Future integration of the microresonators with mid-infrared QCLs could open the path to novel simple and compact mid-infrared comb generators. Furthermore, the choice of the microresonator material is not limited to dielectric crystals. As many semiconductors such as silicon (Si), germanium (Ge) and indium phosphide (InP) also exhibit wide transparency windows in the mid-infrared and third-order nonlinearity, a whole class of on-chip molecular spectrometers based on this approach is conceivable.

\section{Methods}
\textbf{Experimental setup to generate microresonator based mid infrared frequency combs.}
The employed pump laser is the idler beam of a high-power single-frequency, continuous-wave (CW) optical parametric oscillator (OPO) (Aculight Argos Model 2400). The pump of the OPO is a 15~W ytterbium-doped fibre based source operating at 1064~nm, and the OPO is a four mirror ring resonant cavity using a temperature-controlled MgO-PPLN crystal as the nonlinear element. The idler beam is tuneable from $2.4$ to $3.2~\mu$m with up to 3~W output power, corresponding to a signal beam from $1.9$ to $1.6~\mu$m. The idler beam can be fine tuned by piezoelectric tuning of the pump fibre laser. Both the idler and the signal beams have linewidths of less than 1~MHz. 
        The tapered-fibre waveguide is made of low-OH fused silica fibre, which is single-mode at $2.5~\mu$m. The idler beam of the OPO was coupled into the tapered-fibre by an AR-coated aspherical ZnSe lens. Coupling efficiency of more than 60\% was obtained. The optical spectra of the combs were recorded by a Yokogawa long-range optical spectrum analyzer.   
      A second OPO, similar to the pump laser source, is used as the narrow-linewidth CW laser in the beat note measurement. The beat notes are recorded using an extended InGaAs detector with 100 MHz bandwidth.

\section{Acknowledgments}
T. W. H. and N. P. acknowledge support by the European Associated Laboratory "European Laboratory for Frequency Comb Spectroscopy" , the Max Planck Foundation and the Munich Centre for Advanced Photonics. T. J. K. acknowledges funding from the Swiss National Science Foundation, the DARPA QuASAR program and the Max Planck Institute of Quantum Optics. T.J.K. and R. H. acknowledge support by a Marie Curie Industry-Academia Partnerships and Pathways. C. Y. W. acknowledges financial support from a Marie Curie Fellowship (project IRcomb).

\end{document}